\documentstyle[pre,aps,multicol,epsfig]{revtex}
\epsfclipon

\begin{document}   

\tightenlines

\title{Evidence for Slow Velocity Relaxation in Front Propagation in Rayleigh-B\'enard Convection}
\author{Julien Kockelkoren,$^{1,2}$,  Cornelis
Storm$^1$\thanks{Present address: Department of Physics and Astronomy, University of
Pennsylvania, Philadelphia, PA 19104, U.S.A.}, and Wim van Saarloos,$^1$}

\address{
$^1$ Instituut--Lorentz, 
Universiteit Leiden, Postbus 9506, 2300 RA Leiden, the Netherlands\\
$^2$ CEA --- Service de Physique de l'Etat Condens\'e, 
Centre d'Etudes de Saclay, 91191 Gif-sur-Yvette, France}
\maketitle
\begin{abstract}
Recent theoretical work has shown that so-called pulled fronts
propagating into an 
unstable state always converge very slowly to their asymptotic speed
and shape.
In the the light of these predictions,   we reanalyze earlier
experiments by Fineberg and 
Steinberg on  front propagation in a Rayleigh-B\'enard cell. In
contrast
to the original interpretation, we argue that in the experiments the
observed front velocities were some 15\% below the asymptotic front
speed and that this is in rough agreement with the predicted slow
relaxation of the front speed for the time scales probed in the
experiments. We also discuss the possible origin of the unusually
large variation of the wavelength of the pattern generated by the
front as a function of the dimensionless control parameter.
\end{abstract}

\begin{multicols}{2} 

\narrowtext

\section{Introduction}

Although the  propagation of a front into an unstable state 
plays an important role in  
various physical situations ranging from the pearling instability
\cite{moses,powers} to
dielectric breakdown \cite{streamers}, detailed experimental
tests  of the explicit theoretical predications, especially those for
the  velocity of so-called ``pulled'' fronts are scarce.
One of the reasons lies in the difficulty in preparing the system in
the unstable state.

If the initial front profile is steep enough the propagating front
converges to a unique shape and velocity. Theoretically,
one  distinguishes  two regimes for front propagation into unstable
states: the so-called ``pushed'' regime, where 
the front is driven by the nonlinearities and the so-called ``pulled'' regime
where  the asymptotic  velocity of the propagating front, $v_{as}$, equals 
the  spreading speed $v^*$ of linear perturbations around the unstable
state: $v_{as}=v^*$. Pushed fronts are by definition those for which
the asymptotic speed $v_{as}$ is larger than $v^*$: $v_{as}>v^*$.
 It is thus as if a ``pulled''  front is literally 
``pulled'' by the leading edge whose
dynamics is driven by linear instability of the unstable state \cite{dee,benjacob,wvs89,es98}; the
nonlinearities merely cause saturation behind the front. We focus here
on the experimental tests of the dynamics of such pulled fronts; since
$v^*$ is determined by the equations linearized about the unstable
state, the front velocity of pulled fronts can often be calculated
explicitly, even for relatively complicated situations.

There have been two experiments aimed at testing the predictions for
the speed of pulled fronts.  Almost
20 years ago Ahlers and Cannell \cite{ahlers} studied the propagation
of a vortex-front into the laminar state 
 in rotating Taylor-Couette flow. The measured
velocities were about 40\% smaller than expected from the theoretical
predictions. A few years later, however, 
Fineberg and Steinberg (FS) \cite{fs}  published data which appeared
to confirm  the expected velocity in
a Rayleigh-B\'enard convection experiment to within about 1\%. 
The issue then seemed to be settled when it was also shown that the
discrepancy  
observed by Ahlers could be traced back to slow transients
\cite{niklas}.

The theoretical developments of the last few years give every reason
to reconsider the old experiments by FS: It has been shown \cite{es98}
that the convergence of the velocity of pulled fronts is {\em always}
very slow, in fact  with leading and subleading universal terms of
${\mathcal O}(1/t)$ and ${\mathcal O}(1/t^{3/2})$ with prefactors which
follow from the linearized
equation. This slow relaxation implies that it will in general be very
difficult to measure the asymptotic front speed to within a percent or
so in any realistic experiment. Hence, from this new perspective, the
proper question is not why in the Taylor-Couette experiment the
measured velocity was too low, but why apparently in the
Rayleigh-B\'enard experiment of FS the asymptotic front speed was
measured. 

The main purpose of this paper is to address this issue, and 
to reanalyze the experiments in the light of the present theory. We
will conclude that the data of FS actually do show  signs of the
predicted power law 
convergence of the front velocity to an extrapolated asymptotic value which is
about 15\% larger than their transient value. This of course implies
that there is then a discrepancy of order 15\% between the value of
$v_{as}=v^*$ as extrapolated from their data, and the one claimed in
the original experiments. We will argue
that the most likely reconciliation of the two results is that the
value of the correlation length $\xi_0$ in the experimental cell of FS
is somewhat larger than the theoretical value used by FS to interpret
their data. 

Of course, {\em only new experiments can settle whether the interpretation
we propose  is the correct one}. We do consider new experiments along
the lines of FS in fact very desirable, not so much as they might
settle the numerical value 
of the velocity, but more because they hold the promise of being the
first accurate experimental test of the universal power law relaxation
of pulled fronts.

In section II we will first summarize the relevant theoretical
predictions for the velocity of pulled fronts. Then we
will discuss the experiments of FS in the light of these results in
section III, where we will also reanalyze their data. Finally, in
section IV we turn to a brief discussion of the wavenumber of the
pattern selected by the front. Here, the results of FS were
not quite consistent with the predictions for the asymptotic wavenumber from
the Swift-Hohenberg equation. As we shall discuss, the wavelength of
the pattern is affected by various effects which are not easily
controlled, but the most likely interpretation of the data of FS is that they
did not observe the asymptotic wavelength behind the front, but the local 
wavenumber in the leading edge of the front.
Indeed it  is in general difficult to test the theory by studying
the  asymptotic pattern wavelength and the convergence
to the asymptotic value experimentally.

\section{Summary of theoretical predictions} 
\subsection{Asymptotic speed and power law convergence}

Just above the onset of a transition to stationary finite wavelength
patterns, for small dimensionless control parameters $\epsilon$ the
slow dynamics
on length scales  larger
than the  wavelength of the pattern  can be described by the  Ginzburg-Landau amplitude
equation \cite{ch,walgraef}
\begin{equation}
\tau_0 \partial_t A = \epsilon A + \xi_0^2 \partial^2_x A - g |A|^2 A~.
\label{ampeq}
\end{equation}
The time scale $\tau_0$ and length scale $\xi_0$ as well as the
nonlinear saturation parameter $g$ depend on the particular system
under study.

The asymptotic spreading speed $v^*$ of linear perturbations around
the unstable state 
is in general obtained from the linear dispersion relation $\omega(k)$ of a
Fourier mode $e^{-i\omega t +i k x}$ through 
\begin{equation}
\left. \frac{\partial {\rm Im} \omega}{\partial {\rm Im} k} \right|_{k^*} -
v^*=0, ~~
\left. \frac{\partial {\rm Im} \omega}{\partial {\rm Re} k} \right|_{k^*} =0 ,
~~\frac{{\rm Im} \omega(k^*)}{{\rm Im} k} = v^*.
\end{equation}
This yields for the Ginzburg-Landau equation \cite{note3}
\begin{equation}
v^*= 2 \epsilon^{1/2}  \xi_0 \tau_0^{-1}~.\label{v*form}
\end{equation}
and
\begin{eqnarray}
\mu^* &\equiv& \mbox{Im} k^* = \sqrt{\epsilon}/ \xi_0~,\\ D & \equiv &
\frac{1}{2}\left. \frac{ \partial^2 {\rm Im} \omega}{
(\partial {\rm Im} k)^2} \right|_{k^*}   = \frac{\xi^2_0}{\tau_0}~.
\end{eqnarray}
In the experiments with which we will compare, the scaled velocity
\begin{equation}
\tilde{v}= v \frac{\tau_0}{ \xi_0 \sqrt{\epsilon}}
\end{equation}
is often used. According to (\ref{v*form}), for a pulled front the
asymptotic value $\tilde{v}_{as} =2$.

The above results were known in the eighties, at the time when the
experiments were done. The crucial insight of the last few years is the
finding that the convergence or relaxation towards the asymptotic
velocity $v^*$ of pulled fronts is {\em always} extremely slow: the general expression for the time
dependent velocity $v(t)$ emerging from steep initial conditions
(i.e., decaying faster than $e^{-\mu^* x}$) is given by \cite{es98,note3}
\begin{equation}
v(t)=v^* - \frac{3}{2 \mu^* t} + \frac{3}{2 \mu^{*2} t^{3/2}}
\sqrt{\frac{\pi}{D}} +  {\mathcal O}(t^{-2}) ~. \label{relax}
\end{equation}
 For the case of the Ginzburg-Landau amplitude equation, we then
have 
\begin{equation}
\tilde{v}(t)=2 - \frac{3}{2 \epsilon t/\tau_0 } + \frac{3
\sqrt{\pi}}{2  (\epsilon t /\tau_0)^{3/2}} + \cdots \label{asympscaled}
\end{equation}
It is important to stress that  the above expression for $v(t)$ is exact but
asymptotic --- this is illustrated by the fact
that at time $\epsilon t /\tau_0 = \pi$ the subdominant $t^{-3/2}$
term is equal 
to the first correction term of order $t^{-1}$ in absolute value, but
of opposite sign.  Thus, although for any sufficiently long time $t$,
the above expression 
will always become accurate, at any finite time, however, the
expression might only yield a good estimate. In fact, in practice one
usually has
to go to dimensionless times $\epsilon t/ \tau_0 $ of order 10 or larger for
the asymptotic expression to become accurate, while for dimensionless
times $\varepsilon t / \tau_0$ in the range 3 to 10 the first
correction term yields a reasonable order-of-magnitude estimate
\cite{es98}. As we shall 
see below, in the Rayleigh-B\'enard experiments \cite{fs} the maximum
dimensionless time $\epsilon t /\tau_0$ that can be probed is about 3 to
4. In comparing with experiments and in making order of magnitude
estimates, we will therefore  only use the first correction
term.
Finally it is important to realize that
after how long a time
these expressions become really accurate, depends also on the initial
conditions.

\subsection{Dependence on initial conditions}

\begin{figure}
\begin{center}
\epsfig{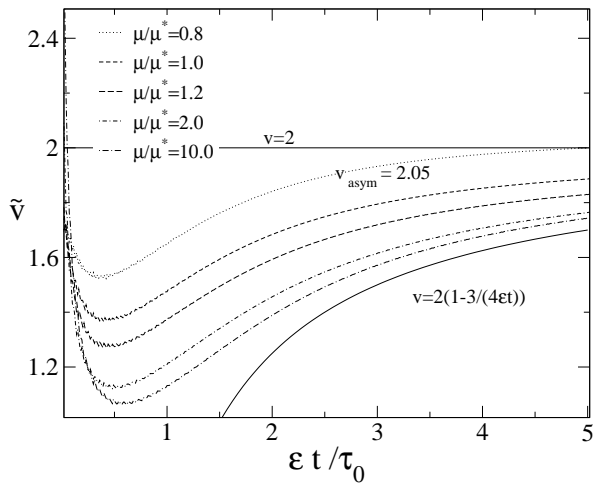}
\end{center}
\caption[]{Velocity of fronts propgating propgating into the unstable
state $A=0$ of the real Ginzburg-Landau equation
with exponentially decaying initial conditions
$A=\sqrt{\epsilon/g}\: e^{-\mu x}$. The velocity is obtained by
tracing the position of the point where $A$ reaches half its
asymptotic value. Since $\epsilon$ and
$\xi_0$ just set the length and time scale, $\tilde{v}(t)$ plotted as
a function of the scaled time $\epsilon t/\tau_0$ is
parameter-independent. Different initial conditions or tracing a
different value of $A$ to determine the velocity yield different
transient behavior, but for sufficiently steep initial conditions, all
curves 
converge to 
the analytic formula for late times.
}
\label{f1}
\end{figure}

In order to illustrate how accurate these expressions are in practice,
we numerically integrate the real Ginzburg-Landau equation starting with an
exponentially decaying initial condition with steepness $\mu$ \cite{note3}:
$A(x,t=0) = a e^{-\mu x}$. The
result is shown  in Fig. \ref{f1} for various values
of $\mu$. We see that initially the velocity falls off
quickly and then approaches the asymptotic velocity from below, and
that the asymptotic 
expression (\ref{asympscaled}) in practice does yield a reasonable
estimate of the time dependent velocity for values of the scaled time
of order 3 and larger. 

We also note that the theoretical analysis shows that for initial
profiles falling off exponentially with steepness $\mu<\mu^*$,
the asymptotic velocity lies above $v^*$ and is given by  $v_{as}(\mu)=
\mu + \frac{1}{\mu}$. The dotted line in Fig. \ref{f1} shows
an example of a case with $\mu$ slightly less than $\mu^*$,
for which $\tilde{v}_{as} = 2.05$. In this case, the time-dependent
scaled velocity is approximately equal to 2 at times of order 3 to
4. 

In the experiments, the typical protocol was to
simultaneously  increase the heat flux from an initial state at
$\epsilon_i <0$ (with a typical value of $\epsilon_i =-0.015$) to a
supercritical value $\epsilon_f$, and switching on the end
heater. A rapid convergence to the asymptotic velocity
due to special initial conditions with $\lambda\approx \lambda^*$
would have required setting $\epsilon_i \approx -\epsilon_f$ for all
$\epsilon_f$   and  would probably also have required switching on  the end heater
before the value of $\epsilon$ was changed, so that effectively at the end a
convection pattern was prepared whose amplitude decayed exponentially
into the bulk. 
Thus, it is theoretically possible to select special initial conditions so as to
get a scaled velocity around 2 at some finite time, but the finetuning
necessary to do so is so sensitive that we consider it  very unlikely that experimental
observations done over a range of values of $\epsilon$ are due to
initial condition effects. As stressed before, only new experiments
can completely rule out this possibility, however.

\section{Reexamination of the  Rayleigh-B\'enard fronts}

In the long quasi-one-dimensional Rayleigh-B\'enard cell of FS \cite{fs}
 the front propagation was initiated by
simultaneously increasing the heat flux to a supercritical value
$\epsilon > 0$ ($\epsilon=\frac{\Delta T}{\Delta T_c} -1$) and
switching on a heater at the ends of the long cell. A vortex front is 
induced near the 
this heated end-wall and the propagation of this front 
into the unstable conductive state is then studied. Thin fins were
attached to the 
long sides of the cell to avoid both induction of long rolls and
pinning. Both because of the fact that  the initial perturbation
was caused by heating at the end and because the state before
bringing the temperature difference beyond its critical value was
unrelated to the final value of $\epsilon$, there is every reason to
believe that the experimental protocol did not  create any
special initial conditions that can not be considered sufficiently
steep or localized.

As FS point out, since the front velocity $v^*$ grows as
$\sqrt{\epsilon}$ while 
the growth rate in the bulk grows as $\epsilon$,  fronts can
only be observed and in fact dominate the dynamics for small enough
$\epsilon$.  
In practice the pattern could be distinguished from the bulk noise up to
a time $ t_{bg} = n \tau_0 \epsilon^{-1}$ where the numerical factor
$n$ is of order 3--4.
This determines some upper limit $\epsilon_0$ on $\epsilon$ for which the front can advance of
the order of $(1/m)$th of the cell length $l$ before bulk growth takes
over:
\begin{equation}
\epsilon_0= \left(\frac{\tilde{v} \xi_0 m n}{l}\right)^2.
\end{equation}

Let us now estimate, using our analytical estimate
(\ref{asympscaled}), the relative importance of the correction term at
the latest times of order $t_{bg} = n\tau_0 \epsilon^{-1} $ at which measurements can be
taken. Substitution gives
\begin{equation}
\tilde{v}(t_{bg})\approx 2 \left( 1 - \frac{3 }{4n} \right) .
\end{equation}
Thus, for the latest time accessible in the experiment,  we obtain
with the empirical experimental value $n=$3--4 a velocity
which  is of order $20 \pm 5 \%$ below the asymptotic value. Although the
asymptotic formula may not be accurate yet at such early times, the
numerical results of Fig. \ref{f1} lead one to expect corrections of
the same order of magnitude: these 
times also correspond to scaled times of order $n$ in the numerical
simulation plots of Fig. \ref{f1}, and as we have seen, over this time
range, the velocity is also suppressed by about 15\% relative to the
asymptotic value. Hence, by any
reasonable estimate, the slow relaxation can not be   negligible in
the experiments!

As stated in the introduction, FS quoted a measured  velocity
$\tilde{v}=2.01 \pm 0.02$ for their experiments, contrary to our
expectation that their scaled velocity should have been in the neighborhood of  1.7. According to the theoretical results, one should expect to get to within 1\% of the asymptotic velocity only around a dimensionless time of order 100!

In our view, the most plausible explanation for
 the origin of this discrepancy is that  the value $\xi_0$ was
actually larger in the experiments than the value used in the
analysis. The value of  $\tau_0$ was experimentally confirmed to be
very close to  the theoretical value  \cite{jay,notetau}. For $\xi_0$,
however, the theoretical value was used without independent
experimental check \cite{jay}. Because of the special design of the cell with
side-fins to create one-dimensional patterns, a different value of
$\xi_0$ might not be unexpected. In fact, an indication that $\xi_0$
in  the experiments was larger than the theoretical value used in the
analysis, comes from the observation of FS that the value of the
wavelength at onset was 13\% larger than the theoretical value. This
might indicate that all lengths in the experiments are a factor 1.13
larger than the theoretical values, and this is precisely the factor
needed to reconcile the front data with the theoretical
expectations! One should keep in mind, though, that $\xi_0$ is
determined by the  the curvature of $\epsilon$ versus $k$ tongue around the
critical wavenumber, and that it is not guaranteed that both are changed by the
same factor; only independent measurements can fully settle this
issue.

We now show that a reanalysis of the data of FS actually gives quite
convincing 
evidence for slow convergence effects in the experimental fronts.

\begin{figure}
\begin{center}
\epsfig{figure=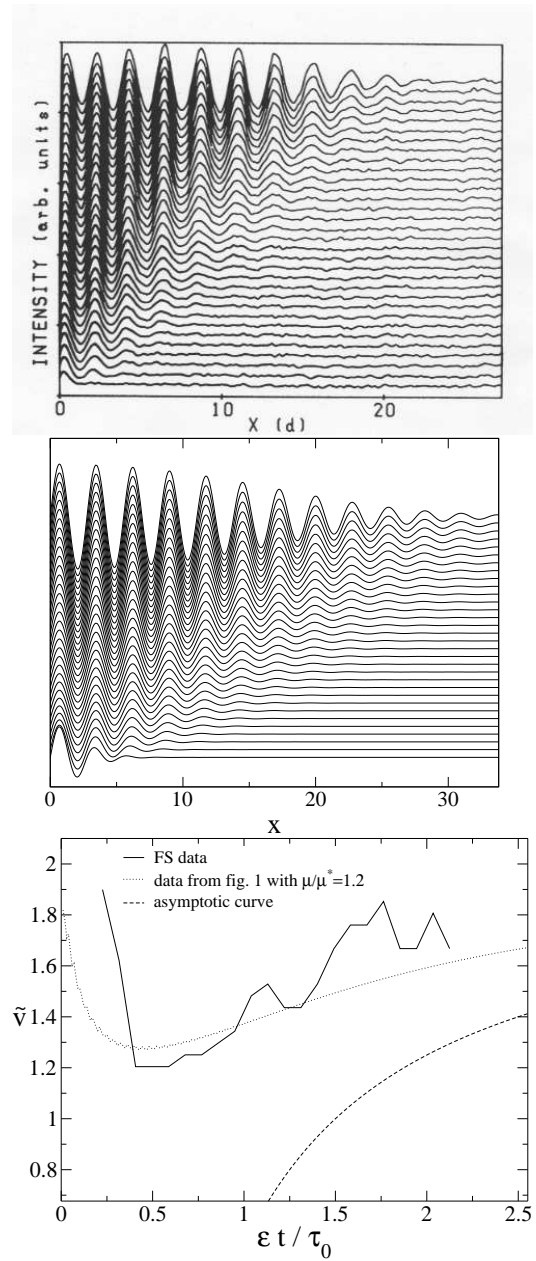,width=0.8\linewidth}
\epsfig{figure=fig2b.eps,width=0.7\linewidth}
\epsfig{figure=fig2c.eps,width=0.8\linewidth}
\end{center}
\caption[]{Top panel: shadowgraph trace of a propagating front in the
experiments of FS  for $\epsilon=0.012$ \cite{jay}. The time difference between
successive traces is 0.42 $t_v$, where $t_v$ is the vertical diffusion
time in the experiments, and the distances are measured in units $d$
(the cell height). (from
\cite{fs}). Middle panel: similar data obtained from numerical integration of the
Swift-Hohenberg equation also at $\epsilon=0.012$ starting with a
localized initial condition. The time difference between successive
traces corresonds to 0.42 $t_v$. 
Bottom panel: Velocity versus time in the experiment, as
obtained by interpolating the maxima of the traces in the top panel, as
explained in the text. The dashed line shows the analytical result
(\ref{asympscaled}) and the dotted curve the result of the amplitude
equation simulation of Fig. \ref{f1} with $\mu/\mu^*
=1.2$. Note that the curves are not fitted, only the absolute scale is
affected by adjusting $\xi_0$.
}
\label{f2} 
\end{figure}

FS measured the velocity by comparing the front with itself
at various time intervals, by appropriately shifting the traces back. This yielded a set of points in the $\Delta
x$, $\Delta t$ plane which appeared to lie on a straight line. However
a possible relaxation of the velocity was masked since points
from early and later times will approximately fall in the same place
in the plane, and because the front shape also has an asymptotic $1/t$
relaxation.  
We therefore have tried to reanalyze the raw data of figure 2 from the
FS paper; this space-time plot of a propagating from is reproduced in
the top panel of our Fig. \ref{f2}. 
We define the position of the front  as the point where the
interpolation of the maxima of the profile equals some fraction of its
maximum in the bulk (we chose 0.4). Our data for $\tilde{v}(t)$
obtained this way are shown in the lower panel of Fig. \ref{f2}. Whereas the local velocity initially slightly decreases, an increase
for dimensionless times larger than 0.5 is evident. We stress that
this qualitative behavior is independent of the choice of the
parameters. In order 
to compare quantitatively to the predictions for the relaxation, we
have used the value for $\tau_0$ given by FS but increased the value
of $\xi_0$ by 13\% on the basis of the argument given above. Clearly,
with this choice, the data are certainly consistent with the analytical as
well as numerical estimates of the velocity relaxation --- in fact,
in a way the data are the first experimental indications for the
universal power law relaxation of pulled fronts \cite{noteahlers}.

\section{Relaxation of wavelength} 
FS also studied the problem of the selection of 
the wavelength
$\lambda$. As we discussed before, the actual value of the wavelength $\lambda_c$
of the patterns is at criticality about 13\% off from the theoretical value;
however, we are not interested here in the absolute value, but in the
relative variation of $\lambda_c/\lambda$. 

The difficulty of comparing theory and experiment on the variation of
the wavelength is that the only theorically sharply defined quantity
is the wavelength sufficiently far behind the front,
$\lambda_{as}$, and that one has to go beyond the lowest order
Ginzburg-Landau treatment to be able to study the pattern wavelength
left behind. E.g., if we use a Swift-Hohenberg equation for a system
with critical wavenumber $k_c$ and bare correlation length $\xi_0$, 
\begin{equation}
\partial_t u = - \frac{(\xi_0 k_c)^2}{4}\left(1 +
\frac{1}{k^2_c}\frac{\partial^2}{\partial x^2} \right)^2 u + \epsilon u - u^3,
\label{SH}
\end{equation}
then 
a node counting argument \cite{dee,wvs89} yields for the
{\em asymptotic} wavelength $\lambda_{as}$ far behind the front
\cite{wvs89}: 
\begin{eqnarray}
\frac{\lambda_c}{\lambda_{as}} & = & \frac{3 \left(3 + \sqrt{1 + 24 \epsilon/(k_c^2\xi_0^2 )}\right)^{3/2}}{8
\left(2+\sqrt{1 + 24 \epsilon/(k_c^2\xi_0^2 )}\right)} \nonumber \\ & \approx & 1
+\epsilon/(2k_c^2 \xi_0^2) ~~~~
(\epsilon \ll 1 )~.
\end{eqnarray}
In the Rayleigh-B\'enard experiments, $k_c\approx 2.75 /d$, where $d$
is the cell height; the
theoretical value is $\xi_0=0.385d$, so if our conjecture that the
value is some 15\% larger is correct, we get $\xi_0\approx 0.44d$. This
then gives 
\begin{equation}
\frac{\lambda_c}{\lambda_{as}} \approx 1+ 0.34 \epsilon. \label{expval}
\end{equation}
As we stressed already above $\lambda_{as}$ is the wavelength far
behind the front; for a propagating pulled front, there is another
important quantity which one can calculate analytically, the
local wavelength $\lambda^*$ measured in the leading edge of the
front. For the Swift-Hohenberg equation, one gets for this
quantity\cite{wvs89,kees} 
\begin{eqnarray}
\frac{\lambda_c}{\lambda^{*}} & = & \sqrt{1 + \frac{1}{4}\left( \sqrt{1
+24\epsilon/(k^2_c\xi_0^2)} -1 \right) } \nonumber
\\
& \approx & 1+ \frac{3 \epsilon}{2 k_c^2 \xi_0^2} ~ ~~~~
(\epsilon \ll 1 )~~~~\nonumber
\\
& \approx & 1+ \epsilon ,
\end{eqnarray}
where in the last line we have used the experimental values.

\begin{figure}
\begin{center}
\epsfig{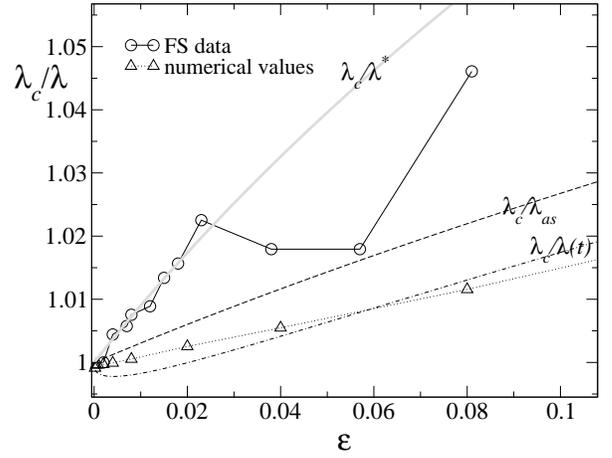}
\end{center}
\caption[]{The selected wave number $\lambda_c/\lambda$ as function of
$\epsilon$. The data points of FS
are denoted by circles, our numerical results on the Swift-Hohenberg
by triangles. The dashed lines shows the prediction for the asymptotic
wavelength, the dashed-dotted line shows the analytic result
with relaxation terms at 
$t=6\tau_0/\sqrt{\epsilon}$ \cite{kees} and the grey line shows the
result for the local wavelength in the leading edge.
}
\label{f3}
\end{figure}

Let us now discuss the experimental findings in the light of these
results. In Fig. \ref{f3} we show both the experimental values of
$\lambda_c /\lambda_{exp}$ as determined experimentally by FS, and
 results for the Swift-Hohenberg equation with the value
(\ref{expval}) relevant for the experiments. 
\\
\hspace*{4mm}{\em (i)} For small $\epsilon$, the experimental values are roughly
linear in $\epsilon$, but when a fit is made over the whole range of
$\epsilon$ values studied, a square root behavior, as proposed by FS, would probably be
better.\\
\hspace*{4mm}{\em (ii)} The experimental values for the wavelength ratio deviate
about a factor of 3 from the theoretically expected value for the
ratio far behind the front,
$\lambda_c/ \lambda_{as}$.\\
\hspace*{4mm}{\em (iii)} Just like the front speed converges very slowly to the
asymptotic value, so does the local value of the wavelength behind the
front \cite{kees}. The relaxation of the velocity is plotted with a
dashed-dotted line in Fig. \ref{f3} for times about
$6\tau_0/\sqrt{\epsilon}$, which is the time it takes for a front to
propagate to close to the center of a system about as large as the
experimental cell. Clearly, the wavelength ratio due to slow
relaxation lies below the asymptotic value, and hence further away
from the experimental data for small $\epsilon$. Also, numerical
results for the wavelength of the fourth ``roll'' measured at the same
time (indicated by the dotted line, lie below the asymptotic curve.\\
\hspace*{4mm}{\em (iv)} FS have measured the wavelength for very small values of
$\epsilon$, down to $4\; 10^{-4}$. However, already for values as small as
0.01, the coherence length $\xi= \xi_0/ \sqrt{\epsilon}$ is about $2
\lambda_c$ in the experiments. The total front width is several times
this number, and for even smaller values of $\epsilon$ the front width
is even smaller. Since the total experimental cell was about $12
\lambda_c$ long in the experiments, it is clear that over much of the
small $\epsilon$ range, one would not expect to see well-developed
fronts in the experiments. In other words, it is very unlikely that
over the experimental range of $\epsilon$-values, one has a chance to
measure $\lambda_{as}$ of a well-developed pattern behind a front.\\
\hspace*{4mm}{\em (v)} The grey line in the plot shows our analytical result for
the wavelength ratio in the leading edge of the front. Clearly, this
line follows the data for small $\epsilon$ quite well. In view also of
point {\em (iv)} above that it will be hard to obtain well-developed
fronts for small $\epsilon$, we propose as a tentative explanation of
the data  that
in the small $\epsilon$ range, one actually measures the emergent roll
pattern associated with the leading edge of a front. Of course, only
new experiments can decide on the validity of this suggestion.\\
\hspace*{4mm}{\em (vi)} We mention that the variation of the wavelength
ratio with $\epsilon$ depends also on the third order derivative term in the expansion
of the dispersion relation around the critical wavenumber. This term
is not modeled correctly in the Swift-Hohenberg equation, but may have
to be taken into account in a full comparison of theory and
experiment. \\
\hspace*{4mm}{\em (vii)} We finally mention that in the experiments there was an
up-down asymmetry in the rolls. We have investigated whether this
could be a source of the discrepancy between the asymptotic wavelength
ratio and the observed one, by studying a Swift-Hohenberg equation
with a symmetry-breaking quadratic term. However, with this term, the
wavelength ratio apears to decrease away from the experimental values.

\section{Conclusion}
It was recently discovered that quite generally pulled fronts  relax very
slowly to 
their asymptotic velocity.  Comparison of the experimental data for
the velocity with
numerical simulations and  analytical estimates give, in our view,
evidence that these experiments provide clear  signs of the presence
of such slow relaxation effects, although the time scales that can be
probed experimentally are too short to test the general power law
relaxation. Theoretically the only
other viable option to reconcile theory with the interpretation
originally proposed \cite{fs} is that somehow special initial
conditions created an initial convection profile with precisely the
right spatial decay into the bulk. As we discussed in section III, in
our view this is an unlikely interpretation, but only new experiments
can settle this issue completely.

While measurements of the wavelength of the pattern generated by a
front are even more difficult to interpret than those of the velocity,
our analysis indicates that in the small-$\epsilon$ regimes a
well-developed front does not fit into the experimental cell, and that
as a result one probes the local wavelength in the leading edge of the
front rather than the well-developed asymptotic wavelength behind it. The analytical estimates are consistent
with this suggestion.

We hope that this work will trigger new experimental activity to
investigate these issues --- experiments along these lines hold the
promise of being the first ones to see the universal power law
relaxation of pulled fronts.

\section{Acknowledgement}
We are grateful to Jay Fineberg and Victor Steinberg for
correspondence about their work and about the issue discussed in this
paper. WvS would also like to thank Guenter Ahlers for urging him to
analyze to what extent slow convergence plays a role in the
Rayleigh-B\'enard experiments. JK is grateful to the `Instituut-Lorentz' for hospitality.

\end{multicols}


\begin{thebibliography}{99}
\bibitem{note3} In \cite{es98} the steepness is denoted by $\lambda$
instead of $\mu$, but we prefer to use $\mu$ here for the
wavelength of the pattern.
\bibitem{notetau} The value of $1/\tau_0$  is given to be 6.90  in
\cite{fs}, but this is a misprint. The theoretical value is 17.94; it
has been explicitly verified that this value accurately describes the
experiments \cite{jay}.
\end{thebibliography}
\end{document}